**The Subtlety of Optimal Paternalism in a Population with Bounded Rationality**

Charles F. Manski
Department of Economics and Institute for Policy Research, Northwestern University, Evanston, IL USA
Corresponding Author: cfmanski@northwestern.edu

and

Eytan Sheshinski
Department of Economics, Hebrew University of Jerusalem, Jerusalem, Israel

Abstract

We study the subtlety of optimal paternalism when a utilitarian planner has the power to design a discrete choice set for a heterogeneous population with bounded rationality. We first consider the planning problem in abstraction. We show that the policy that most effectively constrains or influences choices depends multiplicatively on the preferences of the population and the choice probabilities conditional on preferences that measure the suboptimality of behavior. We then study two settings in which the planner may mandate an action or decentralize decision making. One setting supposes that individuals measure utility with additive random error and maximize mismeasured rather than actual utility. Then optimal planning requires knowledge of the distribution of measurement errors. The other setting studies binary treatment choice when the planner can mandate a treatment conditional on publicly observed personal covariates or can enable individuals to choose their own treatments conditional on private information. Here we focus on situations where bounded rationality takes the form of deviations between subjective and objective probabilities of uncertain outcomes. To illustrate, we consider clinical decision making in medicine. In toto, our cautionary analysis shows that determination of optimal policy requires the planner to possess extensive knowledge that is rarely available. We warn that research in behavioral public economics should avoid overoptimistic claims regarding the nature of optimal paternalistic policies. We argue that credible study of utilitarian planning should consider not only the population but also the planner to be boundedly rational.

Acknowledgements: This co-authored paper developed from solo research initiated in the early 2000s by Sheshinski with the title "Socially Desirable Limits on Individual Choice" and from more recent research by Manski on social planning under uncertainty. Presentations by Sheshinski of his original work were made in various settings, including a seminar at the University of Bonn in 2004 and as the Richard Musgrave Lecture in Munich in April 2012. The research by Manski has been published in journal articles and books. At various stages, we have benefitted from the comments of Peter Diamond, Jacob Goldin, Joram Mayshar, Yeshaya Nussbaum, and Daniel Reck. We have benefitted from the opportunity to present this work in a conference at Duke University in September 2023 and at the Normative Economics and Economic Policy webinar in September 2024.

## 1. Introduction

### 1.1. Background

A central mission of research in public economics has been to determine policies that optimize utilitarian welfare, recognizing that policy choice affects individual behavior. To ease analysis, economists have maintained simplifying assumptions about behavior. It is well known that findings on optimal policy are sensitive to these assumptions.

The seminal Mirrlees (1971) study of optimal income taxation assumed that individuals maximize static deterministic utility when choosing labor supply. It assumed that individuals have homogeneous consumption-leisure preferences and are heterogeneous only in ability, hence wage. Among the assumptions that Mirrlees posed in his introductory section, he stated (p. 176): "The State is supposed to have perfect information about the individuals in the economy, their utilities and, consequently, their actions." His analysis showed that the optimal tax structure is sensitive to the assumed utility function and ability distribution. In his conclusion he wrote (p. 207): "The examples discussed confirm, as one would expect, that the shape of the optimum earned-income tax schedule is rather sensitive to the distribution of skills within the population, and to the income–leisure preferences postulated. Neither is easy to estimate for real economies."

The ensuing literature on optimal taxation has studied settings where individuals may have jointly heterogeneous preferences and abilities, it being assumed that the planner knows the joint distribution of preferences and ability. It has long been recognized that optimal policy is sensitive to the form of this distribution (e.g., Sheshinski, 1972; Atkinson and Stiglitz, 1980). However, empirical understanding of the actual population distribution of preferences and abilities has remained weak, impeding application of the theory. Manski (2014) studied partial identification of income-leisure preferences using revealed-preference analysis of labor supply and reached this pessimistic conclusion (p. 146): "As I see it, we lack the knowledge of preferences necessary to credibly evaluate income tax policies."

Theoretical study of utilitarian policy choice began in the 1700s, was formalized in the first half of the 1900s, and continued to develop steadily through the latter part of the century. The subject has received less attention in the 2000s, as public economists have increasingly performed design-based empirical research rather than welfare-economic study of potential policies, a trend discussed in Manski (2026). A welcome exception to the recent dearth of welfare-economic research is a new body of analysis of optimal paternalistic planning in populations with bounded rationality.

For about two centuries, paternalistic planning was considered outside the domain of utilitarian welfare economics. Economists largely agreed with this emphatic declaration of John Stuart Mill in *On Liberty* (Mill, 1859, p. 18):

> "the only purpose for which power can be rightfully exercised over any member of a civilised community, against his will, is to prevent harm to others. His own good, either physical or moral, is not a sufficient warrant. He cannot rightfully be compelled to do or forbear because it will be better for him to do so, because it will make him happier, because, in the opinions of others, to do so would be wise, or even right."

Paternalistic planning has become a mainstream concern of public economics since the early 2000s, stimulated by the growth of empirical research in behavioral economics in the latter part of the 1900s. Since then, behavioral public economists have suggested that social planners should limit the choice options available to individuals to ones deemed beneficial from a utilitarian perspective or, less drastically, should frame the options in a manner thought to influence choice in a positive way. Thaler and Sunstein (2003) evocatively wrote that such policies express "libertarian paternalism." Their use of informal reasoning to advocate "nudge" policies (Thaler and Sunstein, 2008) has been influential, implemented through organizations such as the Behavioral Insights Team (https://www.bi.team/) initially established by the UK government.

An early expression of the type of formal analysis that we think desirable was given by O'Donoghue and Rabin (2003), who began their article as follows (p. 186):

> "The classical economic approach to policy analysis assumes that people always respond optimally to the costs and benefits of their available choices. A great deal of evidence suggests,

> however, that in some contexts people make errors that lead them not to behave in their own best interests. Economic policy prescriptions might change once we recognize that humans are humanly rational rather than superhumanly rational, and in particular it may be fruitful for economists to study the possible advantages of *paternalistic policies* that help people make better choices.
>
> We propose an approach for studying optimal paternalism that follows naturally from standard assumptions and methods of economic theory: Write down assumptions about the distribution of rational and irrational types of agents, about the available policy instruments, and about the government's information about agents, and then investigate which policies achieve the most efficient outcomes. In other words, economists ought to treat the analysis of optimal paternalism as a mechanism-design problem when some agents might be boundedly rational."

Economists have subsequently performed a growing set of analyses of the type sought by O'Donoghue and Rabin, addressing different classes of policy choices and assuming various distributions of preferences and deviations from rationality. The literature has expanded to the extent that a review article by Bernheim and Taubinsky (2018) is over 100 pages long and contains over 350 references. The literature continues to grow in multiple directions, to the extent that it would be foolhardy for anyone to assert full awareness of all of the current frontiers. Our present paper emerges from the conjunction of two features of the literature to date, one that we find admirable and the other deeply concerning.

We find admirable that much research in behavioral public economics evaluates policy with recognition that individuals have heterogeneous preferences and deviations from rationality. Early on, O'Donoghue and Rabin (2003) referred to "the distribution of rational and irrational types of agents" in their opening paragraphs, quoted above. Camerer *et al.* (2003) called attention to the importance of heterogeneity to policy formation when they recommended adoption of what they termed *asymmetric paternalism*, writing (p. 1212): "a regulation is asymmetrically paternalistic if it creates large benefits for those who make errors, while imposing little or no harm on those who are fully rational." Although some subsequent research work assumed representative agent models for simplicity, study of planning in heterogeneous population is increasingly the norm. See, for example, O'Donoghue and Rabin (2006), Allcott and Taubinsky (2015), Goldin and Lawson (2016), Rees-Jones and Taubinsky (2018), Handel,

Kolstad, and Spinnewijn (2019), Moser and Olea de Souza e Silva (2019), Farhi and Gabaix (2020), Goldin and Reck (2022), Allcott *et al.* (2025), and Lockwood *et al*. (2025).

Our deep concern is that researchers in behavioral economics commonly assert optimal paternalistic planning to be feasible in heterogeneous populations. To their credit, authors sometimes observe that findings on optimal policy are sensitive to assumptions made about the population distribution of preferences and deviations from rationality. They also sometimes caution that the sensitivity of optimal policy to these assumptions makes it important to have a firm empirical understanding of behavior in populations with bounded rationality. Nevertheless, the literature has failed to cope with a severe identification problem that is endemic in empirical research using data on choice behavior to infer the population distribution of preferences and deviations from rationality.

The core of the identification problem was understood by Samuelson (1938, 1948) in his pioneering study of revealed preference analysis, later extended by Afriat (1967). Consider a rational person making decisions with full knowledge of utility outcomes. Observation of a choice reveals only that this action is weakly preferred to all other elements of the choice set. It does not reveal the full preference ordering.

Whereas Samuelson and Afriat contemplated analysis of choices made by one individual, random utility analysis has studied the identification problem at the population level. Marschak (1960), assuming only that preferences are strict with probability one, derived sharp bounds on certain counterfactual choice probabilities. The discrete-choice literature pioneered by McFadden (1974) has mainly assumed for simplicity that the preference distribution lies in a specified finite-dimensional family, yielding point identification of preference distributions. However, being troubled by the limited credibility of parametric models, econometricians have also studied semiparametric and nonparametric models. Whereas point identification of preference distributions persists with modest weakening of parametric models, partial identification becomes the norm when weaker assumptions are maintained. See Manski (2007a, 2014) and Molinari (2020).

As serious as the identification problem is when studying deterministic rational behavior, its severity increases qualitatively when studying boundedly rational behavior. In research performed before development of the behavioral public economics field, several partial identification findings were proved in Manski (1997, 2003). A particularly simple and striking finding reported in Manski (1997), Proposition 1 considered a setting of binary treatment choice and binary utility outcomes. We summarize here:

*Choice Between a Treatment Mandate and Decentralization*: Suppose that a planner can mandate treatment A or B. Alternatively the planner can decentralize decisions, enabling population members to choose their own treatments. Let treatment utility outcomes be binary, taking the value 0 or 1. Assume the planner knows the welfare yielded by each mandate, namely $P[u(A) = 1]$ and $P[u(B) = 1]$. However, the planner does not know the joint distribution of preferences and deviations from rationality in the population.

There are two polar cases. If all population members are rational and maximize utility, decentralized welfare is $P[u(A) =1 \text{ or } u(B = 1]$. If they all irrational and minimize utility, welfare is $P[u(A) = u(B) = 1]$. It is proved that knowledge of welfare under each mandate implies that decentralized welfare can take any values in the interval $[\max(0, C - 1), \min(C, 1)]$, where $C \equiv P[u(A) = 1] + P[u(B) = 1]$. This interval is informative from the left or the right but not from both sides. Its width narrows toward 0 as C approaches 0 or 2 but widens toward 1 as C approaches 1. Thus, knowledge of welfare under the mandates may reveal a lot or a little about decentralized welfare, depending on the empirical value of C. ■

Some subsequent research has revealed more about the identification problem. Assuming that individuals maximize subjective expected utility when making decisions under uncertainty, Manski (2004) called attention to the difficulty of jointly inferring preferences and expectations from choice data. Barseghyan et al. (2021) studied partial identification when bounded rationality takes the form of choice from consideration sets, which are cognitively determined subsets of feasible choice sets. In their review article on behavioral public economics, Bernheim and Taubinsky (2018) wrote (p. 395-96):

> "**Model uncertainty.** The BRP approach is also demanding on analysts because it presupposes that they can successfully identify correct behavioral models. Because behavioral economists operate within a domain that offers abundant degrees of freedom, many distinct models of choice processes

> can potentially account for the same or similar choice mappings. Experience teaches us that building a professional consensus for the "right" model can be extremely difficult, even when the choice mapping is known."

Here "model uncertainty" is close to a synonym for partial identification and "BRP" is an acronym for "behavioral revealed preference."

In light of the above, we think it highly unrealistic for behavioral public economists to assert that they are able to learn optimal policies. Yet this have been the prevailing practice. In the language of Manski (2011, 2020), researchers perform policy analysis with *incredible certitude*.

### 1.2. Contribution of the Paper

This paper argues that, to enhance credibility, behavioral public economists should analyze policy as a problem of social planning under uncertainty, a subject studied in Manski (2024). The paper does much more than only exhort the research community. We use a simple yet broadly applicable framework to increase understanding of the sensitivity of optimal policy to the distribution of population preferences and deviations from rationality. Performing original research, we present a set of novel findings.

We consider a planner who has the power to design a discrete choice set from which individuals will choose. We suppose that there is no social cost to offering larger choice sets. Hence, classical utilitarian welfare economics recommends that the planner should offer the largest choice set possible. We depart from the classical setting by supposing that individuals may be boundedly rational in the sense that they may not choose options that maximize objective expected utility. In such settings, it may be optimal for the planner to constrain the choice set to prevent persons from choosing inferior actions or, less drastically, to frame the choice set in a manner that influences behavior. This paper mainly studies policies that constrain the choice set, particularly mandates.

We use the term *bounded rationality* in the way that Simon (1955) had in mind in the article that spawned the modern literature in behavioral economics (p. 101):

> "Because of the psychological limits of the organism (particularly with respect to computational and predictive ability), actual human rationality-striving can at best be an extremely crude and simplified approximation to the kind of global rationality that is implied, for example, by game-theoretical models."

O'Donoghue and Rabin (2003) similarly wrote: "humans are humanly rational rather than superhumanly rational." Thus, we assume individuals have well-defined stable latent utility function that they want to maximize but find it infeasible to do so. We do not consider more radical conceptualizations of behavior that question the realism of stable utility functions, as in Tversky and Kahneman (1986) and Bernheim and Rangel (2008).

*Example 1*: Some public social security systems and private pensions have an early eligibility age at which a person can start receiving a pension, with less than full benefits. This age differs widely across countries. In the US, partial benefits are obtainable at age 62 and full benefits later (historically at age 65, in process of advancing to 67). The UK has had a single State Pension Age determining eligibility for full benefits (historically at age 65, in process of advancing to 67), with no option of earlier retirement with lower benefits. Imposing a constraint on the earliest age for eligibility hurts workers who would sensibly stop working before this age due to health and other personal circumstances. On the other hand, it prevents people from retiring too early, reflecting shortsightedness. Setting an early eligibility age should strike a balance between these considerations. ■

*Example 2*: Clinical practice guidelines (CPGs) in medicine make treatment recommendations that act as quasi-mandates. Guidelines condition these recommendations on specified publicly observed patient covariates. Clinicians commonly observe patient covariates beyond those considered in guideline recommendations, enabling more refined personalization of treatment. Utilitarian theory assuming that both guideline panels and clinicians act with complete rationality implies that decentralized treatment is preferable to mandates. This conclusion may not hold if guideline panels make optimal recommendations conditioning on the covariates they observe, but clinicians do not. ■

The planner's problem is straightforward if all members of the population have the same known latent preferences. Then the optimal paternalistic utilitarian policy calls on the planner to determine the population-wide best option and mandate it. This obvious result holds regardless of the nature of bounded rationality in the population.

Our concern is settings in which persons have heterogeneous stable preferences and may vary in how their choices deviate from maximization of objective expected utility. Throughout the paper, we assume that members of the population have well-defined latent preferences across actions, expressed as the objective expected utility of actions conditional on privately available information. A person who maximizes objective expected utility will be said to have *complete rationality*.[1]

To begin, Section 2 shows that the policy that most effectively constrains or influences individual choice depends in a particular multiplicative way on the preferences of the population and on the choice probabilities conditional on preferences that measure the suboptimality of behavior. This mathematically simple finding is central to our subsequent analysis. As far as we are aware, the finding is new to research in behavioral public economics. It is, however, reminiscent of a known result on regret in statistical decision theory analysis of treatment choice with sample data on treatment response.

To simplify notation, Sections 2 and 3 suppress publicly observable covariates of members of the population and study optimal planning in a population of persons who are observationally identical to the planner. The findings in Sections 2 and 3 are presented in a set of six simple yet instructive propositions.

---

[1] Applied economists have long associated rationality with maximization of objective expected utility. See, for example, Friedman and Savage (1948). Nevertheless, other decision criteria for choice under uncertainty that may also warrant the term *complete rationality*. For example, maintaining the concept of an objective probability distribution to characterize uncertainty, Manski (1988) studied maximization of objective quantile utility.

A different perspective is manifest in axiomatic decision theory, which considers rationality to mean behavior is consistent with specified choice axioms. Objective probability distributions rarely appear in axiomatic theory. Savage (1954) associated rationality with maximization of subjective expected utility, where subjective probability distributions and utility functions are constructs implied by adherence to specified choice axioms.

Section 2 formalizes how preferences and choice probabilities interact to determine population welfare with each policy. We compare the optimal mandate and decentralized choice, where there is no intervention by the planner. We characterize settings in which the optimal mandate performs better or worse than decentralized choice.

Section 3 considers policy choice when individuals are boundedly rational in a specific way, this being that they measure utility with additive random error and maximize mismeasured rather than actual utility functions. Studying this type of bounded rationality enables more detailed analysis. When the errors in utility measurement are bounded or are independent and identically distributed, we obtain lower bounds on the welfare achieved by decentralized choice and by policies that constrain the choice set. When the errors have a scaled version of the Type I extreme value distribution, implying that choice probabilities have the multinomial logit form, we show that the scale of the error distribution succinctly characterizes the degree of rationality in the population. For any policy that does not mandate a particular treatment, population welfare increases with the degree of rationality, which affects choice probabilities. However, an instructive numerical example shows that the optimal paternalistic policy varies in a subtle way with the degree of rationality. Contrary to what may seem intuitive, the optimal constrained choice set need not monotonically expand as the degree of rationality increases.

In Sections 2 and 3, the distinction between deterministic choice and choice under uncertainty is not relevant, so we refer for brevity to utility rather than to objective expected utility. In Section 4, we address settings with uncertainty and refer explicitly to objective expected utility. We focus on the type of bounded rationality that occurs when subjective probabilistic beliefs of members of the population differ from objective probabilities of uncertain outcomes (aka deviations from rational expectations).

Section 4 removes the simplification that members of the population are observationally identical. We now suppose that members of the population have publicly and privately observable covariates. The planner sees only the publicly observable covariates, whereas members of the population also see the privately observable ones. In these settings, utilitarian theory assuming complete rationality recommends decentralization to enable decision making to condition on private information. However, members of a population with bounded rationality may make sub-optimal decisions.

Considering problems of binary treatment choice under uncertainty, we present the conventional utilitarian argument for decentralization and then focus on a form of bounded rationality that has long been of interest in economics. This is maximization of subjective expected utility when probabilistic beliefs differ from objective probabilities of uncertain outcomes; that is, deviation of beliefs from rational expectations. We characterize the situations in which such deviations yield sufficiently sub-optimal decentralized choices that the optimal mandate improves utilitarian welfare.

The contexts to which our analysis applies range from medical treatment to school tracking to timing of pension eligibility. To illustrate, we focus on medical treatment under uncertainty. Here, the planning entities are panels that formulate clinical practice guidelines, mentioned above in Example 2. We use assessment of women's risk of developing breast cancer to illustrate that patients may condition their beliefs on privately observed covariates not used in guidelines for prophylactic treatment of breast cancer. We call attention to psychological research that studies empirical deviations of clinical judgement from rational expectations.

We then present a formal analysis comparing adherence to CPGs and clinical judgment in medical choice between surveillance and aggressive treatment of disease. This shows that deviations of clinical judgement from rational expectations do not, per se, imply that maximization of subjective expective utility lowers welfare relative to the ideal optimum. Importantly, the distance between subjective and objective probabilities of illness does not determine the welfare performance of patient care with clinical judgement. Considering patients with a specified utility function, what matters are choice probabilities. These are determined by the frequency with which subjective and objective probabilities of illness differ in whether they are smaller or larger than the value of a threshold probability, where the mandate and decentralized choice yield the same objective expected utility.

In Sections 2 through 4, we assume that the planner has the knowledge necessary to choose an optimal policy. Yet we caution throughout that this extensive knowledge is commonly not available. In toto, our analysis characterizes the subtle nature of optimal policy, whose determination requires the planner to possess extensive knowledge of population preferences and behavior that is rarely available.

The concluding Section 5 argues that study of utilitarian policy choice should consider not only the population but also the planner to be boundedly rational. We recommend that behavioral public economists abandon the conventional research practice of determining optimal policy with unsubstantiated assumptions posed for convenience. Instead, researchers should study planning under uncertainty, using credible partial knowledge of population preferences and deviations from rationality.

## 2. The Policy Choice Problem

### 2.1. General Setup and Findings

Let J denote the population of concern to a utilitarian planner. Let C denote a pre-specified largest feasible finite choice set that the planner may make available to each member of J. Let each individual $j \in J$ have a utility function $u_j(\cdot): C \rightarrow R$ expressing the person's preferences. Let $u_j^* \equiv \max_{c \in C} u_j(c)$. This specification of utility functions assumes the absence of social interactions; that is, utility varies with a person's own chosen action but not with those chosen by others.

To formalize utilitarian welfare, consider J to be a probability space with distribution P(j) and let $P[u(\cdot)]$ denote the population distribution of utility functions. Let utility functions be cardinal and interpersonally comparable. Then the idealized optimum utilitarian welfare, if all persons maximize utility, is $E(u^*)$.

Having bounded rationality, individuals may not maximize utility. Let the planner choose among a set S of policies that may constrain or influence choice behavior. Suppose that, with policy s, person j chooses $c_j(s) \in C$. For each $i \in C$, let $P[c(s) = i|u(\cdot)]$ denote the fraction of persons with utility function $u(\cdot)$ who would choose option i under policy s. The utilitarian welfare achieved by this policy is

$$(1) \qquad E\{u[c(s)]\} = \int \sum_{i \in C} u(i) \cdot P[c(s) = i|u(\cdot)]dP[u(\cdot)].$$

The optimal feasible welfare is achieved by a policy that solves the problem $\max_{s \in S} E\{u[c(s)]$.

Observe that the value of $E\{u[c(s)]\}$ depends multiplicatively on the utility functions $u(\cdot)$ and the conditional choice probabilities $P[c(s)|u(\cdot)]$ of the population, averaged across its members. It is revealing to consider the regret of a policy, its degree of sub-optimality, relative to the idealized optimum utilitarian welfare $E(u^*)$. The regret of policy s is

$$(2) \qquad E(u^*) - E\{u[c(s)]\} = \int \sum_{i \in C} [u^* - u(i)]P[c(s) = i|u(\cdot)]dP[u(\cdot)].$$

For each utility function $u(\cdot)$ and action i, $[u^* - u(i)]P[c(s) = i|u(\cdot)]$ is the degree of sub-optimality of i multiplied by its choice probability. Thus, the regret of policy s is a weighted average of the multiplicative interactions of choice probabilities for actions and their degrees of sub-optimality.[2]

In this formalization of bounded rationality, the specific cognitive processes that lead individuals to deviate from utility maximization are immaterial. What matters for social welfare is the set of choice probabilities and the degree of sub-optimality of these choices. This structure of sub-optimality is central to our analysis henceforth.

### 2.1.1. Deterministic and Stochastic Pareto Comparison of Policies

The welfare ranking of two policies, say s and s', generally depends on the distribution of utility in the population. However, the ranking is invariant to $P[u(\cdot)]$ if one policy is Pareto superior to the other in the classical or a weaker sense.

---

[2] This multiplicative structure is similar to the regret of a utilitarian rule for binary treatment choice using sample data on treatment response. There, regret is the mean welfare loss when a member of the population is assigned the inferior treatment, multiplied by the expected fraction of the population assigned this treatment. See Manski (2007a, Section 12.3) and Manski (2021, Section 2.3.2).

Classical Pareto superiority of s relative to s' occurs if $u_j[c_j(s)] \geq u_j[c_j(s')]$ for all $j \in J$ and $u_j[c_j(s)] > u_j[c_j(s')]$ for some $j \in J$. A weaker stochastic sense of Pareto superiority that suffices for utilitarian welfare analysis occurs if $E\{u[c(s)|u(\cdot)]\} \geq E\{u[c(s')|u(\cdot)]]$ for almost every utility function $u(\cdot)$ and $E\{u[c(s)|u(\cdot)]\} > E\{u[c(s')|u(\cdot)]]$ for a set of utility functions that occur with positive probability in the population.

When the choice set contains two actions, say $C = \{a, b\}$, there exists a simple characterization of stochastic Pareto superiority in terms of choice probabilities. It suffices to consider utility functions for which $u(a) \neq u(b)$. Policy s increases the choice probability for the better of two actions if and only if it decreases the choice probability for the worse action. It holds immediately that policy s is stochastically Pareto superior to s' if and only choice probabilities satisfy these inequalities: $P[c(s) = a|u(\cdot)] \geq P[c(s) = b|u(\cdot)]$ for almost all $u(\cdot)$ such that $u(a) > u(b)$, $P[c(s) = b|u(\cdot)] \geq P[c(s) = a|u(\cdot)]$ for almost all $u(\cdot)$ such that $u(b) > u(a)$, and at least one of these inequalities is strict for a set of utility functions that occur with positive probability in the population.

No similarly simple characterization of stochastic Pareto superiority is feasible when C contains more than two actions. In these settings, increasing the choice probability for the best action decreases the sum of choice probabilities for the remaining actions, but it does not necessarily decrease the choice probability for every remaining action. Policy comparison requires attention to the multiplicative interaction of all choice probabilities and the cardinal structure of the associated utilities.

### 2.1.2. The Focus on Choice in Empirical Research in Behavioral Economics

We think it important to point out that empirical research in behavioral economics has commonly focused on choices alone, rather than on the multiplicative interaction of choice probabilities with utilities that determines utilitarian welfare. Thus, behavioral economists have done much more to document the existence of bounded rationality than to measure the severity of its implications.

An apt illustration is the famous Tversky and Kahneman (1981) Asian Disease framing experiment. The authors performed an experiment on decision making that studied how choice behavior depends on the *framing* of the decision problem; that is, on the language that the researcher uses to describe alternative actions and their outcomes. They reported striking results.

We do not question their findings. Choice probabilities varied dramatically with the framing of the decision problem. However, the findings reveal nothing about the magnitudes of the utility losses that subjects would experience if they were to make different decisions. The authors conjectured that the findings support their prospect theory (Kahneman and Tversky, 1979), but, their research provided no direct evidence on the preferences of subjects.

In fact, regret is zero if subjects are risk neutral, a possibility discussed in Manski (2007b, Chapter 15). All of the policy choices posed in the experiment yield the same expected result for survival and death in the population, namely that 200 people will live and 400 will die. Thus, a risk-neutral person is indifferent among the policies. Regret is zero, whatever the choice probabilities may be.

## 2.2. The Optimal Mandate

In the Introduction, we distinguished between policies that constrain and influence individual choices. Both types are encompassed in the above general setup. A choice-constraining policy limits the effective choice set to some $C(s) \subset C$, implying that $P[c(s) = i|u(\cdot)] = 0$ for all $i \notin C(s)$ and all $u(\cdot)$. Such a policy is a mandate if $C(s)$ is a singleton. Mandating action i yields welfare $E[u(i)]$. Hence, the optimal mandate selects an action $i^m$ that solves the problem $\max_{i \in C} E[u(i)]$.

The optimal mandate yields the idealized optimal utilitarian welfare if all members of the population have the same preferences. It yields weakly lower welfare when preferences are heterogeneous. This holds by Jensen's Inequality, which implies that $\max_{i \in C} E[u(i)] \leq E[\max_{i \in C} u(i)]$. The inequality is strict when preferences are sufficiently heterogeneous to make the optimal action vary across the population. The optimal mandate is best for persons who most prefer action $i^m$, but not for those who most prefer other options. By (2), the regret of the optimal mandate is $R_m \equiv E[\max_{i \in C} u(i)] - \max_{i \in C} E[u(i)]$.

Comparison of the optimal mandate with the idealized optimal utilitarian welfare is uninteresting if the population is boundedly rational. Comparisons of interest are to the welfare yielded by feasible policies, whose impacts are determined by the population distribution of utilities and by choice probabilities conditional on utilities. This paper performs various such comparisons.

2.2.1. Comparing the Optimal Mandate with Decentralized Choice

An important comparison, which we perform often in this paper, juxtaposes the optimal mandate with decentralized choice, where the planner does not intervene in private decision making. We denote decentralized choice as the null policy s = o. It yields utilitarian welfare $\int \sum_{i \in C} u(i) \cdot P[c(o) = i|u(\cdot)]dP[u(\cdot)]$. The question of interest is to compare this with the welfare $\max_{i \in C} E[u(i)]$ obtained with the optimal mandate. Decentralized choice provides a heterogeneous population the opportunity to achieve higher welfare than the optimal mandate, but it opens the danger of doing worse. Which prevails depends on the nature of bounded rationality.

A simple baseline result holds if the decentralized choices made by the members of the population are unrelated to their preferences, in the formal sense that choices are statistically independent of utility functions. Proposition 1 proves that the optimal mandate necessarily is preferable to decentralized choice.

*Proposition 1*: Assume that $P[c(o) = i|u(\cdot)] = P[c(o) = i]$ for all $i \in C$ and almost all $u(\cdot)$. Then welfare with the optimal mandate is greater than or equal to welfare with decentralized choice. Welfare with the optimal mandate is strictly greater if there exists $h \in C$ such that $P[c(o) = h] > 0$ and $E[u(h)] < \max_{i \in C} E[u(i)]$. ■

Proof: The assumption implies that

(3) $$\int \sum_{i \in C} u(i) \cdot P[c(o) = i|u(\cdot)]dP[u(\cdot)] = \sum_{i \in C} P[c(o) = i]E[u(i)].$$

$\sum_{i \in C} P[c(o) = i]E[u(i)] \leq \max_{i \in C} E[u(i)]$. The inequality is strict if there exists h s. t. $P[c(o) = h] > 0$ and $E[u(h)] < \max_{i \in C} E[u(i)]$.

Q. E. D.

Although Proposition 1 assumes that choice is unrelated to preferences, it does not assume that all actions have the same choice probability $1/|C|$. If this is additionally assumed, welfare with decentralized choice is the unweighted mean utility across actions, namely $1/|C| \cdot \sum_{i \in C} E[u(i)]$. We henceforth refer to such behavior as *completely random choice*.

Proposition 1 shows that a necessary condition for decentralized choice to outperform the optimal mandate is statistical dependence between choices and utility functions. An important question for policy formation is to characterize the types of dependence that render decentralized choice better or worse than the optimal mandate. This question is difficult to address satisfactorily in abstraction. However, informative results emerge when one considers specific forms of bounded rationality that place structure on choice probabilities. We show one simple result here and others later.

*Utility Maximization with Probability α*

A simple form of bounded rationality with dependence supposes that choices maximize utility with some given probability $\alpha$, and are statistically independent of utility with probability $1 - \alpha$. Then decentralized choice outperforms the optimal mandate if $\alpha$ is sufficiently large and the optimal mandate is better otherwise. Proposition 2 formalizes this result, whose proof is immediate.

*Proposition 2*: Assume that, among persons with each utility function $u(\cdot)$, choice maximizes utility with probability $\alpha$. Choice is statistically independent of utility with probability $1 - \alpha$, with choice probabilities $p_i$, $i \in C$. Then welfare with decentralized choice is $\alpha E(u^*) + (1 - \alpha) \sum_{i \in C} p_i E[u(i)]$. Welfare with the optimal mandate is $\max_{i \in C} E[u(i)]$. Hence, decentralized choice (optimal mandate) outperforms the optimal mandate (decentralized choice) if and only if the former (latter) welfare exceeds the latter (former). ■

Observe that, holding $(p_i, i \in C)$ and $P[u(\cdot)]$ fixed, welfare with decentralized choice increases linearly with $\alpha$. Proposition 2 is consistent with two interpretations of population behavior. In one interpretation, fraction $\alpha$ of the population always maximizes utility and fraction $1 - \alpha$ always behaves in a manner that is unrelated to their preferences. In the other interpretation, each individual temporally varies in rationality, maximizing utility with probability $\alpha$ when facing a given choice problem and experiencing a spontaneous cognitive lapse with probability $1 - \alpha$.

## 2.3. Nudges

A choice-influencing policy enhances the prominence of a specified action but does not prevent persons from choosing other actions. For example, behavioral economists have sought to enhance prominence by specifying some action to be the "default option," by placing it first in the ordering of actions, by associating it with favorable images, and in other ways. Whatever mechanism is used, the objective is to increase the probability with which persons choose this action and decrease the probabilities with which they choose all other actions. Such policies have been called nudges.

A behavioral economist may recommend nudging the population towards the optimal mandate and away from all other alternatives. Let s be such a policy. Comparing it with decentralized choice, the nudge yields these inequalities on choice probabilities: For all $u(\cdot)$, $P[c(s) = i_m|u(\cdot)] \geq P[c(o) = i_m|u(\cdot)]$ and $P[c(s) = i|u(\cdot)] \leq P[c(o) = i|u(\cdot)]$, $i \neq i_m$. It does not seem possible to evaluate the welfare impact of nudges in abstraction. One must consider the context. The remainder of this paper focuses on policies that constrain the choice set and does not study nudges.

## 3. Policy Choice with Additive Error in Utility Measurement

To enable further analysis, we now consider policy choice when individuals are boundedly rational in a specific way. We assume that they measure utility with additive error and maximize mismeasured rather than actual utility functions. We do not assert that this type of bounded rationality is common in actual populations. We study it because the idea is easy to understand and because it enables us to apply findings on choice probabilities developed in the literature analyzing random utility models.

### 3.1. Choice Probabilities Generated by Random Utility Models

Let policy s constrain choice to a choice set C(s), which may be any non-empty subset of C. We assume that, under policy s, person j mismeasures the utility of each $c \in C(s)$ as $u_j(c) + \varepsilon_j(c, s)$, chooses an action $c_j^{\#}(s) \equiv \operatorname{argmax}_{c \in C(s)} u_j(c) + \varepsilon_j(c, s)$, and thus achieves utility $u[c_j^{\#}(s)]$. For simplicity, we assume that the conditional error distribution $P[\varepsilon(c, s), c \in C(s)|u(\cdot)]$ is continuous. This implies that $c_j^{\#}(s)$ is unique for almost every person j. Hence, choice probabilities are well-defined, with

$$(4) \quad P[c^{\#}(s) = i|u(\cdot)] \;=\; P[u(i) + \varepsilon(i, s) \geq u(c) + \varepsilon(c, s), \text{ all } c \in C(s)|u(\cdot)], \quad \text{all } i \in C(s).$$

Inserting these choice probabilities into (1) yields the welfare achieved by policy s, which is

$$(5) \quad E\{u[c(s)]\} \;=\; \int \sum_{i \in C(s)} u(i)\cdot P[u(i) + \varepsilon(i, s) \geq u(c) + \varepsilon(c, s), \text{ all } c \in C(s)|u(\cdot)]dP[u(\cdot)].$$

Other research studying this type of bounded rationality include Allcott and Taubinsky (2015), Goldin and Reck (2022), and Allcott *et al.* (2025).

Equation (4) provides a random-utility model interpretation of bounded rationality (McFadden, 1974; Manski, 1977). The values of the choice probabilities are determined by the conditional error distribution $P[\varepsilon(c, s), c \in C(s)|u(\cdot)]$. In the absence of restrictions on this distribution, any choice probabilities are possible. Hence, assuming that a random utility model expresses bounded rationality does not, per se, yield restrictions on population welfare. Some knowledge of the error distributions is necessary.

For example, Goldin and Reck (2022) used an additive random utility model with a particular type of error distribution to study nudge policies that distinguish some action as the default option. For $i \in C$, let $s_i$ denote a policy specifying i as the default option. Let there exist a person-varying but not action-varying quantity $\gamma_j \geq 0$ such that $\varepsilon_j(i, s_i) = 0$ and $\varepsilon_j(c, s_i) = -\gamma_j$ when $c \neq i$. Thus, an individual subtracts an *as-if cost* $\gamma_j$ from the utility of each action that is not the default. In their analysis, some person-specific fraction of $\gamma_j$ is normative, in the sense that it reduces actual welfare. The remaining fraction is considered to be a mistake, which does not reduce actual welfare. They characterized utilitarian welfare in this setting.

In what follows, we consider random utility models with other assumptions on the error distribution. To our knowledge, this analysis is new to the literature.

### 3.2. Degrees of Rationality

An initial question is whether one may reasonably compare the *degrees of rationality* implied by different error distributions. Our concern is with the population welfare achieved with boundedly rational decision making, so it seems reasonable to say that one error distribution conveys a higher degree of rationality than another if it yields larger mean welfare for the persons who mismeasure utility with it. We do not see a satisfactory way to compare all distributions in this manner. However, comparison is straightforward when the choice set contains two actions and the distributions being compared differ only by a scaling factor. Proposition 3 gives the result:

*Proposition 3*: Let the choice set be $C(s) = \{a, b\}$. Consider any utility function for which $u(b) > u(a)$. Let $P[\varepsilon_0(c, s), c \in \{a, b\}|u(\cdot)]$ be a specified error distribution. Let $\varepsilon(c, s) = \varepsilon_0(c, s)/q(s)$ be $\varepsilon_0$ rescaled by a positive scaling factor $q(s)$. The mean welfare of persons with utility function $u(\cdot)$ increases with $q(s)$. ■

Proof: The choice probability for action b is

(6) $$P[u(b) + \varepsilon_0(b, s)/q(s) \geq u(a) + \varepsilon_0(a, s)/q(s)|u(\cdot)] = P[\varepsilon_0(a, s) - \varepsilon_0(b, s) \leq q(s)[u(b) - u(a)]|u(\cdot)].$$

Given that $u(b) - u(a) > 0$, the probability on the right-hand side weakly increases with q(s). The choice probability for action a commensurately decreases with q(s). The result follows.

Q. E. D.

One may expect that the above finding holds when considering larger choice sets. We have not been able to prove this in generality, but we show in Section 3.5 that it holds when choice probabilities have the multinomial logit form. A difficulty encountered when attempting a general analysis is that, when the choice set contains three or more actions, one can only definitively sign how the choice probabilities for the best and worst actions vary with q(s). The former weakly increases with q(s) and the latter decreases with q(s). Choice probabilities for actions yielding intermediate values of utility may increase or decrease with q(s), depending on the specific error distribution and utility function. This complicates analysis.

Our notation q(s) permits the degree of rationality to vary across policies. We think it plausible that the scale of errors in utility measurement may vary in practice. Consider utility measurement as the size of the choice set C(s) increases. Utility measurement may be computationally costly, as Simon (1955) conjectured. Hence, the accuracy of measurements may tend to decrease with the size of the choice set.

### 3.3. Bounded Measurement Errors

A simple restriction on the error distribution is to assume that it has known bounded support. Such an assumption implies a lower bound on welfare with decentralized choice. Proposition 4 considers the simple case where the support is symmetric about zero.

*Proposition 4*: Assume (4). Assume that, for a known $\delta > 0$, $P[-\delta \leq \varepsilon(c, o) \leq \delta | u(\cdot)] = 1$ for all $c \in C$ and almost all $u(\cdot)$. Then $E\{u[c(o)]\} \geq E(u^*) - 2\delta$. Decentralized choice necessarily outperforms the optimal mandate if $E(u^*) - 2\delta > \max_{i \in C} E[u(i)]$. ■

Proof: For each individual j, let $c_j$ be an action that maximizes actual utility on C and let $c_j^{\#}(o)$ maximize mismeasured utility. Given the bounded support assumption, $u_j^* - u[c_j^{\#}(o)] \leq 2\delta$ for almost all $j \in J$. Hence, the first result follows from (2). The second holds as $\max_{i \in C} E[u(i)]$ is the welfare of the optimal mandate.

Q. E. D.

Observe that the lower bound on welfare with decentralized choice increases as $\delta$ decreases. Hence, it is reasonable to say that $\delta$ measures the maximum deviation of the population from rationality.

### 3.4. Simple Scalability

A different lower bound on welfare with decentralized choice emerges if, for almost every utility function $u(\cdot)$, the error components $\varepsilon(c, s)$, $c \in C(s)$ are known to be independent and identically distributed (i. i. d.). We do not assume knowledge of the specific distribution, nor that the distribution has bounded support. Indeed, it may vary arbitrarily with $u(\cdot)$. The i. i. d. assumption expresses the idea that individuals make "white-noise" errors in utility measurement. The error distribution may vary across persons j and policies s. We assume only that errors do not vary systematically across actions.

Manski (1975) showed that, when errors are conditionally i. i. d., choice probabilities are related to utility functions by a set of inequalities called *simple scalability*. For each utility function $u(\cdot)$ and action pair $\{a, b\} \in C(s) \times C(s)$, simple scalability holds if

$$(7a) \quad u(a) > u(b) \Leftrightarrow P[u(a) + \varepsilon(a, s) \geq u(c) + \varepsilon(c, s), \text{ all } c \in C(s)|u(\cdot)]$$
$$\geq P[u(b) + \varepsilon(b, s) \geq u(c) + \varepsilon(c, s), \text{ all } c \in C(s)|u(\cdot)],$$

$$(7b) \quad u(a) = u(b) \Leftrightarrow P[u(a) + \varepsilon(a, s) \geq u(c) + \varepsilon(c, s), \text{ all } c \in C(s)|u(\cdot)]$$
$$= P[u(b) + \varepsilon(b, s) \geq u(c) + \varepsilon(c, s), \text{ all } c \in C(s)|u(\cdot)].$$

Let $u^{mean}(s)$ denote unweighted mean utility in set C(s); that is, $u^{mean}(s) \equiv [1/|C(s)|] \sum_{c \in C(s)} u(c)$. Simple scalability implies this lower bound on welfare with choice from C(s):

*Proposition 5*: Assume that, for almost all $u(\cdot)$, $\varepsilon(c, s)$, $c \in C(s)$ are i. i. d. conditional on $u(\cdot)$. Then $\int u^{mean}(s)dP[u(\cdot)] \leq E\{u[c(s)]\}$. ■

Proof: The assumption implies (7a)-(7b) for almost all $u(\cdot)$. These inequalities show that the ordering of choice probabilities coincides with the ordering of utility values. It follows that the choice-probability weighted average of utility in set C(s) is greater than or equal to the unweighted mean utility. That is,

$$(8) \quad u^{mean}(s) \leq \sum_{i \in C(s)} u(i) \cdot P[u(i) + \varepsilon(i, s) \geq u(c) + \varepsilon(c, s), \text{ all } c \in C(s)|u(\cdot)].$$

The result follows by integrating each side of (8) over the distribution of utilities, and applying (1).

Q. E. D.

When applied to decentralized choice (s = o), the inequality in Proposition 5 is $[1/|C|] \sum_{i \in C} E[u(i)] \leq E\{u[c(o)]\}$. The lower bound on the left-hand side is weakly less than $\max_{i \in C} E[u(i)]$. Hence, Proposition 5 does not imply that decentralized choice outperforms the optimal mandate. It does imply that population welfare is no smaller than welfare with completely random choice.

The assumption that $\varepsilon(c, s)$, $c \in C(s)$ are i. i. d. conditional on $u(\cdot)$ suffices to imply simple scalability but is not necessary. A much weaker assumption suffices if the choice set C(s) contains only two actions, say {a, b}. Then (7a)-(7b) reduce to

(9a) $u(a) > u(b) \Leftrightarrow P[u(a) + \varepsilon(a, s) \geq u(b) + \varepsilon(b, s)|u(\cdot)] \geq \frac{1}{2}$.

(9b) $u(a) = u(b) \Leftrightarrow P[u(a) + \varepsilon(a, s) \geq u(b) + \varepsilon(b, s)|u(\cdot)] = \frac{1}{2}$.

These conditions hold if $P[\varepsilon(a, s) - \varepsilon(b, s) \geq 0|u(\cdot)] = \frac{1}{2}$.

### 3.5. Multinomial Logit Choice Probabilities[3]

A substantial strengthening of the knowledge used above assumes that errors are i. i. d. with a scaled version of the standard type 1 extreme-value distribution, also called the Gumbel distribution. Let $\varepsilon_0(c, s)$, $c \in C(s)$ be independent and have the common distribution function $P[\varepsilon_0(c, s) \leq t] = \exp(-e^{-t})$. Let $\varepsilon(c, s) = \varepsilon_0(c, s)/q(s)$ be $\varepsilon_0$ rescaled by a positive scaling factor q(s). Then the conditional choice probabilities have the multinomial logit form (McFadden, 1974):

[3] The analysis in Sections 3.5 and 3.6 originated in work initiated in the early 2000s that was presented in several seminars and distributed in slides (Sheshinski, 2012), but not circulated as a working paper.

(10) $P[c^{\#}(s) = i|u(\cdot)] \;=\; P[u(i) + \varepsilon(i, s) \geq u(c) + \varepsilon(c, s), \text{ all } c \in C(s)|u(\cdot)] \;=\; e^{q(s)u(i)}/\sum_{c \in C(s)} e^{q(s)u(c)}, \;\; i \in C(s).$

Hence, the welfare achieved by policy s is

(11) $E\{u[c(s)]\} \;=\; \int \sum_{i \in C(s)} u(i)\cdot \left[e^{q(s)u(i)}/\sum_{c \in C(s)} e^{q(s)u(c)}\right] dP[u(\cdot)].$

As in Section 3.2, q(s) quantifies the degree of rationality in the population, under policy s. As $q(s) \to \infty$, the choice probability for the action that maximizes utility increases to one. As $q(s) \to 0$ the choice probabilities for all actions converge to $1/|C(s)|$, as they would be with completely random choice. Proposition 6 shows that population welfare increases with q(s).

*Proposition 6*: Let the choice set be C(s). Assume that conditional choice probabilities have form (10). Then population welfare weakly increases with q(s). Welfare strictly increases with q(s) if a positive fraction of the population have utility functions in which utility varies across the actions in C(s). ∎

Proof: For utility function $u(\cdot)$ and action $i \in C(s)$, the partial derivative of the choice probability with respect to q(s) is

(12) $\partial P[c^{\#}(s) = i|u(\cdot)]/\partial q(s) \;=\; P[c^{\#}(s) = i|u(\cdot)]\{u(i) - v[s, u(\cdot)]\},$

where $v[s, u(\cdot)] \equiv \sum_{h \in C(s)} u(h)\cdot P[c^{\#}(s) = h|u(\cdot)]$ is the choice-probability-weighted expected utility for persons with utility function $u(\cdot)$. Thus, increasing q(s) raises (lowers) the choice probabilities of actions whose utility is greater (smaller) than expected utility.

It follows from (12) that, for persons with utility function $u(\cdot)$, the partial derivative of welfare with respect to $q(s)$ is

$$(13)\quad \partial[\sum_{i \in C(s)} u(i)\cdot \left[e^{q(s)u(i)}/\sum_{c \in C(s)} e^{q(s)u(c)}\right]/\partial q(s)$$

$$= \sum_{i \in C(s)} u(i)\cdot P[c^{\#}(s) = i|u(\cdot)]\{u(i) - \sum_{h \in C(s)} u(h)\cdot P[c^{\#}(s) = h|u(\cdot)]\}$$

$$= \{\sum_{i \in C(s)} u(i)^2\cdot P[c^{\#}(s) = i|u(\cdot)]\} - \{\sum_{h \in C(s)} u(h)\cdot P[c^{\#}(s) = h|u(\cdot)]\}^2.$$

The final expression is the variance of $u[c(s)]$, which must be non-negative. All choice probabilities are positive. Hence, $V\{u[c(s)\} > 0$ if $u(\cdot)$ varies across the actions in $C(s)$.

Q. E. D.

Although multinomial logit choice probabilities have a simple form, study of the variation in population welfare across policies is subtle. To demonstrate, we compare the decentralized-choice policy $s = o$ with another, labelled $s = 1$, that eliminates one action from C, say d. Suppose for simplicity that the degree of rationality is the same for these two policies; that is, $q(o) = q(1)$. Then population welfare with policy $s = o$ may be larger or smaller than with $s = 1$.

Welfare with $s = o$ is smaller than welfare with $s = 1$ for persons whose utility function makes d the action that minimizes utility. For such persons, elimination of d reduces its choice probability to zero and increases the choice probabilities for all remaining actions, which have higher utility. Symmetrically, welfare with $s = o$ is larger than welfare with $s = 1$ for persons whose utility function makes d the action that maximizes utility. Elimination of d may decrease or increase welfare for persons whose utility function makes d an action with intermediate utility. It follows that the overall welfare ranking of the two policies depends on the population distribution of utility and on the degree of rationality.

3.6. Numerical Example Showing the Subtlety of Optimal Choice-Constraining Policy

In Section 3.5, policy s was described by two factors, the set C(s) constraining individual choice and the degree of rationality q(s) with the policy. We now specialize further, considering a set S of policies that yield the same value of q(s), now labeled q, and that differ only in their choice-constraining sets C(s). A policy cannot exclude every option in C, so there exist $2^{|C|} - 1$ such policies.

The welfare yielded by policy s is

$$(14) \quad E\{u[c(s)]\} = \int \sum_{i \in C(s)} u(i) \cdot \left[e^{q \cdot u(i)} / \sum_{c \in C(s)} e^{q \cdot u(c)}\right] dP[u(\cdot)].$$

Analytical determination of a policy that maximizes welfare does not seem feasible, but numerical calculation of welfare is possible given a specification of q and P[u(·)].

A numerical example demonstrates that optimal policy choice is subtle. The example is based on the famous Hotelling (1929) model of choice when individuals and stores are located on a line. Let C contain three actions (potential stores), each identified by a location $x_i$, i = 1, 2, 3 on a line. Let J contain three individuals, each residing at a location $\theta_j$, j = 1, 2, 3 on this line. Let the utility of action i to person j be $u_i(\theta_j) = -(x_i - \theta_j)^2$. Thus, due to transportation costs, utility decreases with the distance of individual j's location, $\theta_j$, from store $x_i$. By construction, preferences are *single-peaked*.

In our example, we specify $x_1 = 0.5$, $x_2 = 1$, $x_3 = 1.6$ and $\theta_1 = -0.5$, $\theta_2 = 1$, $\theta_3 = 2$. This yields the utility values shown in Figure 1:

Figure 1

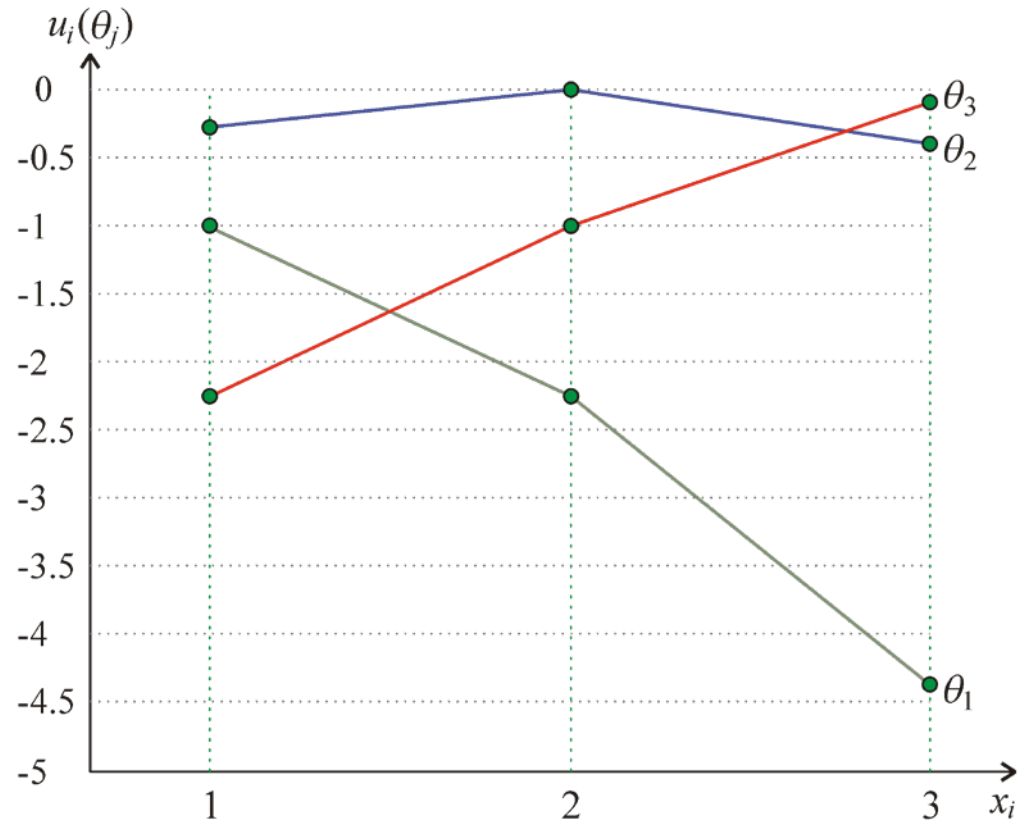

The seven possible constrained choice sets are {1}, {2}, {3}, {1, 2}, {2, 3}, {1, 3}, and {1, 2, 3}. The corresponding social welfare functions are denoted $W^1$, $W^2$, $W^3$, $W^{1,2}$, $W^{2,3}$, $W^{1,3}$, and $W^{1,2,3}$, respectively. Figure 2 plots each of these welfare functions against different integer levels of q. For each q, the optimum choice-set corresponds to *the outer envelope* of these plots.

Figure 2

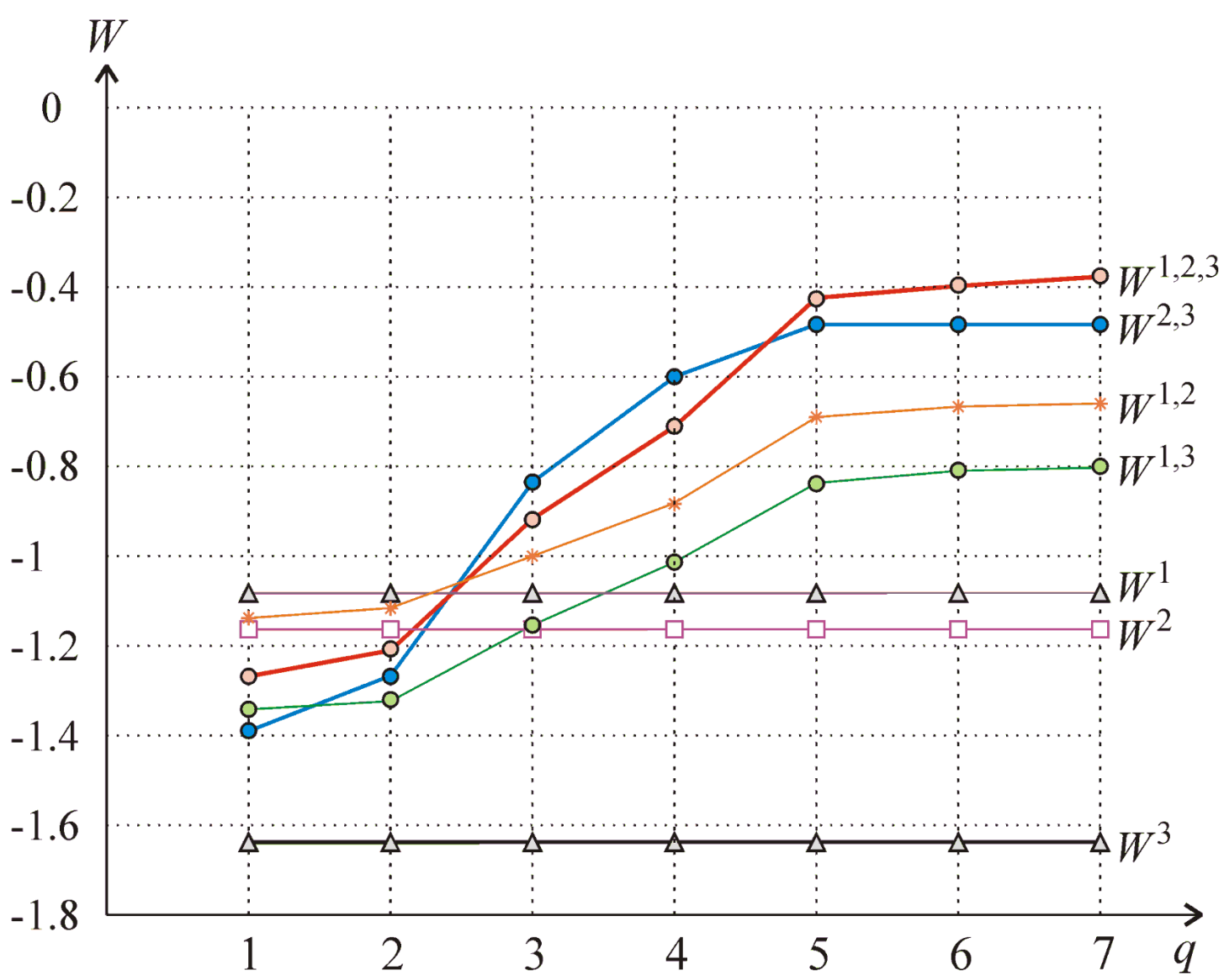

Observe that the optimal constrained choice set has a single action ($W^1$) at low values of q and includes all actions ($W^{1,2,3}$) at high values of q. Of particular interest is the fact that the ordering of welfare across choice sets is not nested, with reswitching occurring as q rises. For example, choice set {1, 2} outperforms set {2, 3} when q is smaller than about 2.5, but the welfare ordering reverses when q is larger. Choice set {1, 2, 3} outperforms set {2, 3} when q is smaller than about 2.2, the welfare ordering reverses for q between 2.2 and about 4.8, and then reverses again for q above 4.8.

This is only an example, but it suffices to demonstrate quantitatively the subtlety of optimal policy choice. We find that, even in a highly simplified environment assuming multinomial logit choice probabilities, the welfare ordering of different choice-constraining policies is rather sensitive to the degree of rationality in the population.

## 4. Mandated or Decentralized Treatment of a Population with Publicly and Privately Observed Covariates

Whereas members of the population were observationally identical to the planner in Sections 2 and 3, we now suppose that each person has publicly observable covariates $x \in X$ and privately observable covariates $z \in Z$, where X and Z are finite sets. We analyze settings in which a planner can mandate particular actions or can decentralize decision making. The planner sees only the publicly observable covariates. Individuals also see the privately observable ones. Thus, the planner can condition a mandate on x but not on z. Decentralized choices can vary with (x, z). The term *complete rationality* now means maximization of objective expected utility conditional on (x, z).

There are many contexts in which a planner chooses between an x-varying mandate and (x, z)-varying decentralized choice. A government may mandate that eligibility for a public pension begins at a particular age or may enable workers to receive a smaller benefit at a younger age if they wish. A school principal may mandate that high school students with covariates x enroll in a mathematics class taught at a specified intellectual level or can permit students to self-select course levels. A clinical guideline panel can recommend a particular medical treatment for patients with covariates x, or the panel can state that physicians who observe these and other patient covariates should use clinical judgement to choose treatments.

Observation of (x, z) expands the set of feasible treatment choices relative to observation of x alone. Utilitarian theory assuming complete rationality recommends decentralization, enabling exploitation of the larger choice set. See, for example, Phelps and Mushlin (1988), Basu and Meltzer (2007), and Manski (2007b). The increase in welfare achieved by observation of (x, z) relative to x is sometimes called the *value-of-information*.[4]

We study a simple yet nuanced setting of binary treatment under uncertainty. If the planner and decentralized decision makers make objectively correct probabilistic predictions of an uncertain utility-relevant outcome, utilitarian theory implies that decentralized treatment outperforms a mandate. We quantify the value of information, paraphrasing analysis in Manski, Mullahy, and Venkataramani (2023).

A mandate may yield higher utilitarian welfare if some decentralized decision makers have bounded rationality and make sub-optimal decisions. Section 4.1 presents the analysis. Section 4.2 considers medical treatment.

### 4.1. Analysis

Let there be two feasible treatments, labeled A and B. Treatment choice is made without knowing a utility-relevant binary outcome, y = 1 or 0. For example, if treatments are mathematics courses taught at different levels, we may have y = 1 if a student would pass the more difficult course and y = 0 if the student would not pass the course. If A and B are options for patient care, y = 1 may mean that a patient is healthy and y = 0 if the patient has an illness of concern.

Each person has covariates (x, z), with x observable by the planner and (x, z) by decentralized decision makers. Let $p_x = p(y = 1|x)$ and $p_{xz} = p(y = 1|x, z)$ be objective probabilities that y = 1 conditional on x and on (x, z). Let each value of z occur for a positive fraction of persons; thus, $P(z|x) > 0$ for all $z \in Z$. Assume that, conditional on x, $p_{xz}$ varies with z.

[4] The present discussion concerns maximization of objective expected utility. Mathematically related but conceptually distinct work in Bayesian statistical decision theory studies the value of information for individual maximization of subjective expected utility. See, for example, Good (1967) and and Kadane, Shervish, and Seidenfeld (2008).

Let $U_x(y, t)$ denote the objective expected utility that a person with covariates x would experience with treatment t, should the outcome be y. For simplicity, this specification assumes that, conditional on x, expected utility does not vary across persons with different values of z. However, z still matters to decision making because the outcome probabilities $p_{xz}$ do vary with z.[5] We assume that the planner knows $U_x(y, t)$ for each possible value of (x, t, y).

Maximum utilitarian welfare using $p_{xz}$ to predict y is always at least as large as using $p_x$, and it is strictly larger if optimal treatment choice varies with z. Manski, Mullahy, and Venkataramani (2023) derived a simple expression that quantifies the value of information. Section 4.1.1 summarizes the derivation. This result provides the foundation for novel consideration of planning with bounded rationality in Sections 4.1.2 and 4.1.3.

4.1.1. Optimal Treatment with Complete Rationality

For any policy s, utilitarian welfare $E\{u[c(s)]\}$ is a group-size weighted average of welfare conditional on x; that is, $E\{u[c(s)]\} = \sum_{x \in X} E\{u[c(s)]|x\} \cdot P(x)$. A planner who observes x and has the authority to vary policy with x maximizes welfare by separately solving the problems $\max_{s \in S} E\{u[c(s)|x]$, $x \in X$. We consider x-specific policies that either mandate a specified treatment or permit decentralized treatment choice.

The optimal x-specific mandate by a utilitarian planner is

(15a) choose treatment A if $p_x \cdot U_x(1, A) + (1 - p_x) \cdot U_x(0, A) \geq p_x \cdot U_x(1, B) + (1 - p_x) \cdot U_x(0, B)$,

[5] For example, in a medical treatment setting, z may be features of an individual's immune system, encoded in DNA. It may be that utility functions do not vary with these genetic features, but risk of illness does vary with them.

(15b) choose treatment B if $p_x \cdot U_x(1, A) + (1 - p_x) \cdot U_x(0, A) \leq p_x \cdot U_x(1, B) + (1 - p_x) \cdot U_x(0, B)$.

With (x, z) privately observable, the optimal decentralized treatment is

(16a) choose treatment A if $p_{xz} \cdot U_x(1, A) + (1 - p_{xz}) \cdot U_x(0, A) \geq p_{xz} \cdot U_x(1, B) + (1 - p_{xz}) \cdot U_x(0, B)$,

(16b) choose treatment B if $p_{xz} \cdot U_x(1, A) + (1 - p_{xz}) \cdot U_x(0, A) \leq p_{xz} \cdot U_x(1, B) + (1 - p_{xz}) \cdot U_x(0, B)$.

With criterion (15), the maximized welfare for persons with covariates x is

$$(17) \qquad \max\, [p_x \cdot U_x(1, A) + (1 - p_x) \cdot U_x(0, A),\ \ p_x \cdot U_x(1, B) + (1 - p_x) \cdot U_x(0, B)].$$

With criterion (16), the maximized welfare for patients with covariates (x, z) is

$$(18) \qquad \max\, [p_{xz} \cdot U_x(1, A) + (1 - p_{xz}) \cdot U_x(0, A),\ \ p_{xz} \cdot U_x(1, B) + (1 - p_{xz}) \cdot U_x(0, B)].$$

In the latter case, the maximized welfare for persons with covariates x is the mean of (18) with respect to the distribution P(z|x); that is,

$$(19) \quad E_{z|x} \left\{ \max\, [p_{xz} \cdot U_x(1, A) + (1 - p_{xz}) \cdot U_x(0, A),\ \ p_{xz} \cdot U_x(1, B) + (1 - p_{xz}) \cdot U_x(0, B)] \right\}.$$

Jensen's inequality shows that, conditional on x, maximum welfare using $p_{zx}$ to predict y is at least as great as maximum welfare using $p_x$. Hence, decentralized decision making outperforms mandating a treatment. However, Jensen's inequality does not quantify the extent to which criterion (16) outperforms (15). Manski, Mullahy, and Venkataramani (2023) do this through direct comparison of the criteria.

Without loss of generality, let the optimal mandate be treatment A; that is, A is optimal in (15). Let A be optimal in (16) for all $z \in Z_A$ and let $Z_B$ be the complement of $Z_A$. Thus, inequality (16a) holds for $z \in$ $Z_A$, some subset of Z, and does not hold for $z \in Z_B$. Criterion (16) yields better outcomes than (15) for persons with $z \in Z_B$ and the same outcomes as (15) for persons with $z \in Z_A$.

Use the decomposition of Z into $(Z_A, Z_B)$ to rewrite (17) and (19) as

$$(20)\quad \max\, [p_x \cdot U_x(1, A) + (1 - p_x) \cdot U_x(0, A),\ \ p_x \cdot U_x(1, B) + (1 - p_x) \cdot U_x(0, B)]$$

$$= p_x \cdot U_x(1, A) + (1 - p_x) \cdot U_x(0, A)$$

$$= P(z \in Z_A|x) \cdot E[p_{xz} \cdot U_x(1, A) + (1 - p_{xz}) \cdot U_x(0, A)]|x, z \in Z_A]$$

$$+\ P(z \in Z_B|x) \cdot E[p_{xz} \cdot U_x(1, A) + (1 - p_{xz}) \cdot U_x(0, A)]|x, z \in Z_B].$$

and

$$(21)\quad E_{z|x}\, \{\max\, [p_{xz} \cdot U_x(1, A) + (1 - p_{xz}) \cdot U_x(0, A),\ \ p_{xz} \cdot U_x(1, B) + (1 - p_{xz}) \cdot U_x(0, B)]\}$$

$$= P(z \in Z_A|x) \cdot E[p_{xz} \cdot U_x(1, A) + (1 - p_{xz}) \cdot U_x(0, A)]|x, z \in Z_A]$$

$$+\ P(z \in Z_B|x) \cdot E[p_{xz} \cdot U_x(1, B) + (1 - p_{xz}) \cdot U_x(0, B)]|x, z \in Z_B].$$

Subtracting (20) from (21) yields

$$(22)\quad P(z \in Z_B|x) \cdot E\{[p_{xz} \cdot U_x(1, B) + (1 - p_{xz}) \cdot U_x(0, B)] - [p_{xz} \cdot U_x(1, A) + (1 - p_{xz}) \cdot U_x(0, A)]|x, z \in Z_B\}.$$

The strict inequality $p_{xz} \cdot U_x(1, B) + (1 - p_{xz}) \cdot U_x(0, B) > p_{xz} \cdot U_x(1, A) + (1 - p_{xz}) \cdot U_x(0, A)$ holds for all $z \in Z_B$. Hence, (22) is positive if $P(z \in Z_B|x) > 0$. We thus find that, with complete rationality, a mandate cannot yield higher population welfare than decentralized treatment. Decentralization outperforms a mandate if the optimal treatment varies across z.

This qualitative finding repeats one obtainable using Jensen's inequality. What is new here is that (22) quantifies the extent to which criterion (16) outperforms (15). The magnitude of (22) is the product of two factors. One is the fraction $P(z \in Z_B|x)$ of persons for whom treatment B yields strictly larger expected utility than treatment A. The other is the mean gain in expected utility that criterion (16) yields for the subset $Z_B$ of persons.

4.1.2. Optimal Treatment with Bounded Rationality

As above, suppose without loss of generality that the optimal mandate for persons with covariates x is treatment A. Again, let A be optimal for $z \in Z_A$ and let $Z_B$ be the complement of $Z_A$. However, suppose that some decentralized decision makers, having bounded rationality, do not use criterion (16) to choose treatments. They choose the worse treatment rather than the better one.

For persons with covariate values (x, z), let $q_{xz}$ denote the choice probability for the better treatment and let $1 - q_{xz}$ be the choice probability for the worse treatment. Then $q_{xz} = 1$ if everyone has complete rationality and $q_{xz} < 1$ if some decision makers have bounded rationality and choose sub-optimally.

In this setting, mandating A continues to yield welfare (20). However, decentralization does not yield (21). Instead, decentralization yields this lower welfare:

$$(23)\quad P(z \in Z_A|x)\cdot E\{q_{xz}[p_{xz}\cdot U_x(1, A) + (1 - p_{xz})\cdot U_x(0, A)] + (1 - q_{xz})[p_{xz}\cdot U_x(1, B) + (1 - p_{xz})\cdot U_x(0, B)]\ |x, z \in Z_A\}$$
$$+\ P(z \in Z_B|x)\cdot E\{q_{xz}[p_{xz}\cdot U_x(1, B) + (1 - p_{xz})\cdot U_x(0, B)] + (1 - q_{xz})[p_{xz}\cdot U_x(1, A) + (1 - p_{xz})\cdot U_x(0, A)]\ |x, z \in Z_B\}.$$

Here, as in Sections 2 and 3, the best policy depends on the multiplicative interaction of population preferences and choice probabilities. Permitting decision makers with covariates x to make decentralized decisions yields higher welfare than mandating treatment A if (23) exceeds (20). The mandate is preferable if (20) exceeds (23). Ceteris paribus, the mandate is better if the choice probabilities ($q_{xz}$ , $z \in Z$) for the optimal z-specific treatments are sufficiently small.

4.1.3. Optimal Treatment When Decentralized Decisions Maximize Subjective Expected Utility

Among the many forms that bounded rationality may take, one that has long been of interest in economics has been maximization of subjective expected utility when probabilistic beliefs differ from objective probabilities of uncertain outcomes. The axiomatic decision theory of Savage (1954) does not refer to an objective reality and, hence, does not view differences between subjective and objective probabilities as bounded rationality. The Savage theory only concerns *procedural rationality,* meaning adherence to certain consistency axioms that related decisions across choice sets. In contrast, our $p_{xz}$ is the objective probability of an uncertain outcome and we use the term *complete rationality* to mean maximization of objective expected utility. From this perspective, maximization of subjective rather than objective expected utility is a form of bounded rationality.

There is much reason to think that deviations from rational expectations are common. Manski (2004) argues that individuals attempting to learn about the real world face identification problems and statistical imprecision in data analysis akin to those that economists face in empirical research. Thus, persons with covariates (x, z) may not know $p_{xz}$.

Let person j place subjective probability $\pi_j$ on the event y = 1 and choose a treatment that maximizes subjective expected utility. Thus, j acts as follows:

(24a) choose treatment A if $\pi_j \cdot U_x(1, A) + (1 - \pi_j) \cdot U_x(0, A) \geq \pi_j \cdot U_x(1, B) + (1 - \pi_j) \cdot U_x(0, B)$,

(24b) choose treatment B if $\pi_j \cdot U_x(1, A) + (1 - \pi_j) \cdot U_x(0, A) \leq \pi_j \cdot U_x(1, B) + (1 - \pi_j) \cdot U_x(0, B)$.

When $\pi_j \neq p_{xz}$, person j maximizes mismeasured expected utility. Criterion (24) does not express mismeasurement as an additive error in the manner of Section 3, but it can equivalently be stated that way.

Maximization of subjective expected utility does not imply that person j behaves sub-optimally. Let $t_x(p_{xz})$ and $t_x(\pi_j)$ denote the treatments solving (16) and (24). Behavior is optimal if $t_x(\pi_j) = t_x(p_{xz})$ and sub-optimal otherwise.

Subjective probabilities may vary across the population. Let $P(\pi|x, z)$ be the objective distribution of $\pi$ across persons with covariates (x, z). The choice probability for the objectively optimal treatment is

$$(25) \quad q_{xz} = P[\pi: t_x(\pi) = t_x(p_{xz})|x, z].$$

Hence, mean welfare conditional on x when treatment is decentralized and decision makers maximize subjective expected utility is given by (23), with choice probabilities (25).

The x-specific welfare yielded by a mandate and by decentralized treatment are functions of the covariate distribution $P(z|x)$ and the values of $[q_{xz}, p_{xz}, U_x(\cdot, \cdot)]$, $z \in Z$. A planner need not have complete knowledge of these quantities to optimally choose between a mandate and decentralization. However, implementation of optimal paternalism requires sufficient knowledge to determine which option yields higher welfare.

### 4.2. Application to Medical Treatment

#### 4.2.1. Clinical Practice Guidelines

Medical textbooks and training have long offered clinicians guidance in patient care. Such guidance has become institutionalized through issuance of clinical practice guidelines (CPGs). Institute of Medicine (2011) writes (p. 4): “Clinical practice guidelines are statements that include recommendations intended to optimize patient care that are informed by a systematic review of evidence and an assessment of the benefits and harms of alternative care options.”

Recommendations made in CPGs are not legal mandates, but clinicians have strong incentives to comply, making adherence close to compulsory. A patient’s health insurance plan may require adherence to a CPG as a condition for reimbursement of the cost of treatment. Adherence may furnish evidence of due diligence that legally defends a clinician in the event of a malpractice claim. Adherence to guidelines provides a rationale for care decisions that might otherwise be questioned by patients, colleagues, or employers.

The analysis in Section 4.1 shows that adherence to a CPG cannot outperform decentralized care if guideline panels and clinicians are utilitarian and have complete rationality. If a CPG conditions its recommendations on all of the patient covariates that clinicians observe, it can do no better than reproduce clinical decisions. CPGs typically condition recommendations on a subset of the clinically observable covariates. Hence, as shown in Section 4.1.1, adhering to a CPG may yield inferior welfare because the guideline does not personalize patient care to the extent possible.

#### 4.2.2. Illustration: Assessment of Illness Risk in CPGs for Prophylactic Care of Breast Cancer

CPGs commonly bring to bear evidence-based objective probabilistic predictions of illness that condition on a subset of the patient covariates that clinicians observe. An apt illustration are guidelines for prophylactic care of women at risk of developing breast cancer. In this setting, option A is routine surveillance, usually meaning that a woman receives a breast examination and mammogram annually or biannually, depending on age. Option B may be some form of enhanced surveillance, such as more frequent mammograms. This more aggressive option does not affect the risk of disease development, but it may reduce the severity of disease outcomes by enabling earlier diagnosis and treatment of tumors. A potential side effect may be an increased risk of cancer caused by the radiation from mammograms.[6]

The utilitarian analysis in Section 4.1.1 shows that, ceteris paribus, some form of aggressive treatment is the better option if the risk of breast cancer is sufficiently high, and routine surveillance is better otherwise. Some CPGs use the Breast Cancer Risk Assessment (BCRA) Tool of the National Care Institute (2024) to assess risk and recommend aggressive treatment if the predicted probability of invasive cancer in the next five years is above a specified threshold. In particular, National Comprehensive Cancer Network (2024) recommends routine surveillance if the predicted probability using the BCRA tool is below 0.017 and some form of aggressive treatment if the probability is higher.

---

[6] Other more aggressive options may include strategies for reduction of the risk of disease development. These include changes to diet, administration of a drug such as tamoxifen, and preventive mastectomy. Each strategy may have side effects, most obviously in the case of preventive mastectomy.

The BCRA Tool gives an evidence-based objective probability that a woman will develop breast cancer conditional on eight attributes: (1) history of breast cancer or chest radiation therapy for Hodgkin Lymphoma (yes/no); (2) presence of a BRCA mutation or diagnosis of a genetic syndrome associated with risk of breast cancer (yes/no/unknown); (3) current age, in years; (4) age of first menstrual period (7-11, 12-13, ≥ 14, unknown); (5) age of first live birth of a child (no births, < 20, 20-24, 25-29, ≥ 30, unknown); (6) number of first-degree female relatives with breast cancer (0, 1, >1, unknown); (7) number of breast biopsies (0, 1, > 1, unknown); (8) race/ethnicity (White, African American, Hispanic, Asian American, American Indian or Alaskan Native, unknown).

The reason that the BCRA Tool assesses risk conditional on these covariates and not others is that it uses a modified version of the "Gail Model," based on the empirical research of Gail *et al.* (1989). The Gail *et al.* article estimated probabilities of breast cancer for white women who have annual breast examinations, conditional on attributes (1) through (7). Scientists at the National Cancer Institute later modified the model to predict invasive cancer within a wider population of women.

The BCRA Tool personalizes predicted risk of breast cancer in many respects, but it does not condition on further patient covariates that may be associated with risk of cancer and that may be observed in clinical practice. When considering the number of first-degree relatives with breast cancer (attribute 6), the Tool does not consider the number and ages of a woman's first-degree relatives, nor the ages when any of them developed breast cancer. These factors should be informative when interpreting the response to the item. Nor does it condition on the prevalence of breast cancer among second-degree relatives, a consideration that figures in another risk assessment model due to Claus, Risch, and Thompson (1994). When considering race/ethnicity (attribute 8), the BCRA Tool groups all white woman together and does not distinguish subgroups such as Ashkenazi Jews, who are thought to have considerably higher risk of a BRCA mutation than other white subgroups, a potentially important matter when the answer regarding attribute (2) is "unknown." Moreover, the BCRA Tool does not condition on behavioral attributes such as excessive drinking of alcohol, which has been associated with increased risk of breast cancer (Singletary and Gapstur, 2001).

#### 4.2.3. Psychological Research Comparing Statistical Prediction and Clinical Judgment

Even though clinicians can usually personalize care beyond the capability of CPGs, the medical literature contains many commentaries exhorting clinicians to adhere to guidelines, arguing that CPG developers have superior knowledge of treatment response than do clinicians.[7] Thus, a rationale for endorsement of CPGs by the medical establishment is a perception that guideline panels make approximately objective probabilistic predictions of health outcomes conditional on publicly observed patient covariates x. In contrast, there is widespread concern that clinicians may form inaccurate subjective beliefs conditional on the covariates (x, z) that they observe. A body of psychological research provide some foundation for this concern. We summarize here, paraphrasing discussion in Manski (2018, 2019).

Psychological research comparing evidence-based statistical predictions with ones made by clinical judgment has concluded that the former consistently outperforms the latter when the predictions are made using the same patient covariates. The gap in performance has been found to persist even when clinical judgment uses additional covariates as predictors. This research began in the mid-twentieth century, notable early contributions including Sarbin (1943, 1944), Meehl (1954), and Goldberg (1968). To describe the conclusions of the literature, we quote the informative review article of Dawes, Faust, and Meehl (1989), who distinguished statistical/actuarial prediction and clinical judgment as follows (p. 1668):

> "In the clinical method the decision-maker combines or processes information in her or her head. In the actuarial or statistical method the human judge is eliminated and conclusions rest solely on empirically established relations between data and the condition or event of interest."

[7] Institute of Medicine (2011) states (p. 26): "Trustworthy CPGs have the potential to reduce inappropriate practice variation." Institute of Medicine (2013) states (p. 2-15): "geographic variation in spending is considered inappropriate or 'unacceptable' when it is caused by or results in ineffective use of treatments, as by provider failure to adhere to established clinical practice guidelines." These and many similar quotations exemplify a widespread belief that adherence to guidelines is socially preferable to decentralized patient care.

Comparing the two in circumstances where a clinician observes patient covariates that are not utilized in available actuarial prediction, they stated (p. 1670):

> "Might the clinician attain superiority if given an informational edge? For example, suppose the clinician lacks an actuarial formula for interpreting certain interview results and must choose between an impression based on both interview and test scores and a contrary actuarial interpretation based on only the test scores. The research addressing this question has yielded consistent results . . . . Even when given an information edge, the clinical judge still fails to surpass the actuarial method; in fact, access to additional information often does nothing to close the gap between the two methods."

Here and elsewhere, Dawes, Faust, and Meehl (1989) cautioned against use of clinical judgment to subjectively predict disease risk or treatment response conditional on patient covariates that are not utilized in evidence-based assessment tools or research reports. They attributed the weak performance of clinical judgment to clinician failure to adequately grasp the logic of the prediction problem. Psychological research published after Dawes, Faust, and Meehl (1989) has largely corroborated the conclusions reached there, albeit occasionally with caveats. See, for example, Groves *et al.* (2000).

The psychology literature challenges the realism of assuming that clinicians have rational expectations. However, it does not per se imply that adherence to CPGs outperforms decentralized decision making using clinical judgment. One issue is that the psychology literature has not addressed all welfare-relevant aspects of clinical decisions. We showed in Section 4.1.1 how optimal decisions are determined by objective expected utilities. Psychologists have only compared medical risk assessments made by statistical predictors and by clinicians. They have not compared evaluations of expected utility, which depends on patient preferences as well as on risk assessments.

A second issue is that psychological research has seldom examined the accuracy of probabilistic risk assessments. It has been more common to assess point predictions of illness. Study of the logical relationship between probabilistic and point predictions shows that data on the latter only yields wide bounds on the former. Suppose that a clinician uses a symmetric loss function to translate a probabilistic risk assessment into a yes/no point prediction that a patient will develop a disease. Then observation that the forecaster states "yes" or "no" as a prediction only implies that he judges the probability of illness to be in the interval [½, 1] or [0, ½] respectively (Manski, 1990). Thus, analysis of the accuracy of point predictions of illness does not reveal much about the accuracy of statistical and clinical assessment of illness probabilities.

Given these and other issues, psychological research does not suffice to conclude that mandating adherence to CPGs is superior to decentralized decision making. Adherence to CPGs may be inferior to the extent that CPGs condition on fewer patient covariates than do clinicians. It may be superior to the extent that imperfect subjective clinical judgment generates sub-optimal clinical decisions. How these opposing forces interplay depends on the setting.

#### 4.2.4. Using CPGs or Clinical Judgment to Choose Between Surveillance and Aggressive Treatment

Consideration of choice between surveillance and aggressive treatment of patients at risk of disease yields insight into the circumstances when adherence to a CPG outperforms decentralized care using clinical judgment. This choice requires resolution of tension between benefits and costs. Aggressive treatment may be more beneficial to the extent that it reduces the risk of disease development or the severity of disease. It may be more harmful to the extent that toxicity generates health adverse events and/or treatment has financial costs beyond those associated with surveillance.

One example of the choice problem is prophylactic care of breast cancer, discussed above. Other prominent cases are the choice between surveillance and drug treatment for patients at risk of heart disease or diabetes. Yet others are choice between surveillance and adjuvant chemotherapy or immunotherapy for patients who have had surgical removal of a cancerous tumor and are at risk of metastasis.

Manski (2018, 2019) characterized the choice between surveillance and aggressive treatment as a subclass of the binary treatment problems studied in Section 4.1, in which the illness-specific expected utilities $u_x(\cdot, \cdot)$ satisfy two inequalities. Recalling the (x, z)-specific decision criterion (16), consider the value of $p_{xz}$ that equalizes the expected utility of the two treatments. This *threshold probability* is

$$(26) \qquad p^*_{xz} = \frac{u_x(0, A) - u_x(0, B)}{[u_x(0, A) - u_x(0, B)] + [u_x(1, B) - u_x(1, A)]} .$$

Let A and B denote surveillance and aggressive treatment. It commonly is credible to suppose that surveillance yields higher expected utility when a patient does not develop the disease, and aggressive treatment yields higher utility when a patient does develop the disease. Formally, assume that

$$(27) \quad u_x(0, A) > u_x(0, B) \text{ and } u_x(1, B) > u_x(1, A).$$

These inequalities imply that $0 < p^*_{xz} < 1$. Moreover, they imply that A is optimal if $p_{xz} \leq p^*_{xz}$ and B is optimal if $p_{xz} \geq p^*_{xz}$.

Consider decentralized patient care for patient j when the clinician does not know the objective probability $p_{xz}$ of illness and maximizes subjective expected utility using subjective probability $\pi_j$. The choice is A if $\pi_j \leq p^*_{xz}$ and B if $\pi_j \geq p^*_{xz}$. This choice is optimal if $\text{sgn}(\pi_j - p^*_{xz}) = \text{sgn}(p_{xz} - p^*_{xz})$ and is sub-optimal if $\text{sgn}(\pi_j - p^*_{xz}) \neq \text{sgn}(p_{xz} - p^*_{xz})$. Considering all patients with covariates (x, z), the objective probability that clinical judgement yields the optimal treatment is

$$(28) \quad q_{xz} = P[\pi\text{: } \text{sgn}(\pi_j - p^*_{xz}) = \text{sgn}(p_{xz} - p^*_{xz}) | x, z].$$

Equation (28) shows that deviations of clinical judgement from rational expectations do not, per se, imply that clinical maximization of subjective expective utility lowers welfare relative to the ideal optimum.

Moreover, the distance between subjective and objective probabilities of illness does not determine the welfare performance of patient care with clinical judgement.

For a given utility function $u_x(\cdot, \cdot)$, what matters are choice probabilities and the losses from making sub-optimal choices, which multiply to yield regret. Choice probabilities are determined by the frequency with which subjective and objective probabilities differ in whether they are smaller or larger than the value of the threshold probability. Losses from sub-optimal choices are determined by the difference between the objective illness probability $p_{xz}$ and the threshold probability $p^*_{xz}$, where treatments A and B have the same objective expected utility.

## 5. Policy Choice When the Planner has Bounded Rationality

We have shown that the optimal utilitarian policy in a heterogeneous population with bounded rationality is highly context specific. The central reason, shown in Section 2, is that the regret of a policy is a subtle multiplicative function of individual utilities and of choice probabilities conditional on utility, summed across the population. The analysis in Sections 3 and 4 shows that determination of an optimal policy requires the planner to have considerable knowledge of population preferences and deviations from rationality, knowledge that is rarely available.

Some research in behavioral public economics has recognized the sensitivity of optimal paternalism to the context, but has optimistically conjectured that empirical analysis can provide planners with the necessary knowledge. In their mainly theoretical study of optimal taxation with boundedly rational agents, Farhi and Gabaix (2019) wrote (p. 313):

> "A difficulty confronting all behavioral policy approaches is a form of the Lucas critique: how do the underlying biases change with policy? The empirical evidence is limited, but we try to bring it to bear when we discuss the endogeneity of attention to taxes . . . . We hope that more empirical evidence on this will become available as the field of behavioral public finance develops."

In their study of optimal default policies, Goldin and Reck (2022) wrote (p. 27):

> "We have shown how the degree to which decision makers make privately suboptimal choices affects optimal default policy, but empirically estimating such internalities is widely acknowledged to be one of the central challenges in behavioral public economics."

They later added (p. 32): "Uncertainty over the decision-making model that generates an observed behavior is a pervasive source of difficulty in behavioral economics." Nevertheless, they retained enough confidence in the power of revealed-preference analysis that they presented an empirical study of optimal default policy in the context of 401(k) plan contribution decisions.

The review article of DellaVigna (2018) on "Structural Behavioral Economics" exemplifies the optimistic perspective of researchers. DellaVigna argues that behavioral economists should estimate parametric structural econometric models of choice behavior of the type developed from the 1970s onwards. He favorably assesses the ability of such models to deliver point estimates of population distributions of utility functions and behavior, enabling design of optimal utilitarian policies. He recognizes that these estimates rest on maintained modeling assumptions, stating (p. 616):

> "A second issue with structural estimation is that the estimates, and ensuing welfare and policy implications, are only as good as the joint set of assumptions going into the model. The estimates may be sensitive to changing some of the auxiliary assumptions, and it is often difficult to thoroughly test the robustness of the estimates. . . . Relatedly, it is easy, after all the work of estimation, to take the welfare and policy implications too much at face value."

Yet he remains optimistic that estimates of point-identified parametric structural models can be sufficiently realistic to provide a suitable basis for conventional public economics study of optimal policy choice.[8]

In contrast to the prevalent perspective in behavioral public economics, we are pessimistic about the feasibility of credible implementation of optimal paternalism. As discussed in the Introduction, even when heterogeneous agents are completely rational, revealed preference analysis requires unrealistically strong

---

[8] Throughout the article, DellaVigna's discussion of model identification maintains the classical binary differentiation between parameters that are identified or not identified. The potential use of credible assumptions to estimate partially identified structural models is not discussed.

In their separate review article in the same volume, Bernheim and Taubinsky (2018) obliquely recognized the phenomenon of partial identification in their brief discussion of "model uncertainty." Nevertheless, they did not address the implications for policy analysis. Their article repeatedly uses the term "optimal policy" and does not engage the difficult problem of social planning under uncertainty.

assumptions to point-identify population distributions of utility functions. Study of partial identification with weaker assumptions shows that credible assumptions commonly yield a large identification region for the distribution of utility functions (Manski, 2007a, 2014).

As mentioned in the Introduction, the identification problem is yet more severe when decision makers may be boundedly rational. Then empirical study of choices must attempt to interpret behavior without the benefit of the transparent linkage of choices to utility rankings available in classical revealed preference analysis. The difficulty is evident when bounded rationality takes the form of maximization of mismeasured utility, as we showed in Section 3 and 4. Then choices reveal the ranking of mismeasured utilities rather than the ranking of the actual utilities that determine welfare.

It follows that utilitarian planners with incomplete knowledge of population preferences and deviations from complete rationality should not seek to optimize policy invoking assumptions that lack credibility. Doing so constitutes policy analysis with incredible certitude. Planners should recognize their own bounded rationality, stemming from incompleteness of their knowledge of population preferences and deviations from rationality.

Abandoning the unrealistic objective of optimization does not nihilistically imply that study of behavioral public economics is fruitless. When realistic optimization is infeasible, planners may make social decisions using various reasonable criteria for planning under uncertainty. Prominent criteria with well-understood properties include maximization of subjective expected social welfare, maximin planning, and minimax-regret planning. Manski (2024) presents broad ideas and a set of applications, mainly studying minimax-regret planning. These applications do not study planning in populations with bounded rationality, but such planning problems may be addressed in the same manner. Study of reasonable utilitarian planning under uncertainty is challenging but can be rewarding. We recommend that behavioral public economics should begin to face the challenge.